\documentclass[showpacs,preprintnumbers,amsmath,amssymb, preprint]{revtex4}
\usepackage{epsf}
\usepackage{graphicx}
\usepackage{dcolumn}
\usepackage{bm}
\usepackage{amsmath}
\usepackage{amsfonts}
\usepackage{amssymb}
\input epsf

\begin{document}

\title{Transport Length Scales in Disordered Graphene-based Materials: \\
Strong Localization Regimes and Dimensionality Effects}

\author{Aur\'elien Lherbier$^{1,3}$, Blanca Biel$^2$, Yann-Michel Niquet$^3$ and Stephan Roche$^4$}

\affiliation{ 
$^1$ Laboratoire des Technologies de la Microélectronique (LTM), UMR 5129 CNRS, CEA 17 Rue des Martyrs 38054 Grenoble France \\
$^2$ CEA, LETI-MINATEC, 17 rue des Martyrs, 38054 Grenoble, Cedex 9 France\\
$^3$ CEA, Institut des Nanosciences et Cryog\'enie, SP2M/L\_Sim,  17 rue des Martyrs, 38054 Grenoble Cedex 9, France\\
$^4$ CEA, Institut des Nanosciences et Cryog\'enie, SPSMS/GT,  17 rue des Martyrs, 38054 Grenoble Cedex 9, France
}

\date{\today}

\begin{abstract}
We report on a numerical study of quantum transport in 
disordered two dimensional graphene and graphene nanoribbons.
By using the Kubo and the Landauer approaches, transport length scales in the diffusive (mean free path, charge mobilities) and localized regimes (localization lengths) are computed, assuming a short range disorder (Anderson-type). In agreement with localization scaling theory, the electronic systems are found to undergo a conventional Anderson localization in the zero temperature limit. Localization lengths in weakly disordered ribbons are found to differ by two orders of magnitude depending on their edge symmetry, but always remain several orders of magnitude smaller than those computed for 2D graphene for the same disorder strength. This pinpoints the role of transport dimensionality and edge effects.\end{abstract}

\maketitle

Recently, single graphene sheet could be isolated 
either from chemical exfoliation of bulk 
graphite \cite{Graphene1}, or by epitaxial 
growth on metal substrates through thermal decomposition of SiC \cite{Graphene2}. 
These technological achievements have opened 
unprecedented opportunities to explore quantum transport 
in low dimensional carbon-based disordered systems \cite{Graphene3,Graphene4}.

Because of the unique electronic properties of the 
2D graphene (massless Dirac fermions with linear 
dispersion and electron-hole symmetry), 
disorder effects and transport properties turn out 
to be unconventional. Theoretically, it has been shown 
that for long range impurity potentials, 
intervalley $K \to K'$ scattering between the two Dirac nodes
could be strongly reduced, resulting in anomalously 
low backscattering rates \cite{AndoJPSJ98}, 
extremely large elastic mean free paths and vanishingly 
small localization effects \cite{WLocalization1}. 
In contrast, for short range impurity potentials
(where all types of scattering 
between $K$ and $K'$ are allowed),
stronger quantum interferences could develop, 
leading to weak localization, or strong Anderson 
localization in the zero-temperature limit \cite{SLocalization2}. 
To date magnetotransport experiments either performed on exfoliated or epitaxial graphene have reported 
both weak antilocalization and weak localization effects 
\cite{WL-WALBerger}, confirming the sensitivity of 2D transport 
in graphene to the external random potential, whose precise origin remains unknown. 

Beyond 2D graphene physics, the transport properties of quasi-1D graphene 
nanoribbons (GNRs) with width down to a few tens of 
nanometers have been characterized \cite{KimIBM}. In contrast to 2D graphene, the electronic properties of GNRs are strongly dependent on confinement effects and edge symmetries \cite{Ribbonpure}. 
These new structures share similarities with carbon nanotubes, 
often viewed as rolled single graphene ribbons, and that have provided unique materials for investigating 1D transport phenomena such as Luttinger liquid and Kondo physics or Anderson localization \cite{RMP}. 

The issue of localization in graphene-based materials 
is currently highly debated from a theoretical standpoint. For instance, the measurement of a minimal conductivity 
of $\sim 2-5 e^{2}/h$ in samples for which
charge mobilities change, however, by almost one order of magnitude 
remains to be fully understood \cite{Graphene1,Graphene2,Graphene3}. 
Indeed, a conventional treatment of disorder effects within the 
self consistent Born approximation (SCBA) yields $\sigma_{xx}^{\rm min}\sim 4e^{2}/(h\pi)$ 
for the two Dirac nodes \cite{AndoJPSJ98} ($h$ the Planck constant), 
hence typically smaller by a $1/\pi$ factor with respect to the experimental data. 
Depending on the disorder model, the use of the Kubo approach suggest 
several scenarii to understand such discrepancies \cite{Kubo}. 
Besides, the role played by quantum interferences and 
the transition to a localization regime in graphene and GNRs remain 
poorly explored but fiercely debated \cite{SLocalization2}.

In this Letter, by using both the Kubo and Landauer approaches, 
the transport length scales in 2D graphene are investigated and 
compared with those of the quasi-1D GNRs. The disorder (Anderson-type) is introduced via random fluctuations of the 
onsite energies of the $\pi$-orbitals, which mimics a short range 
scattering potential that has been widely studied in the past as 
a generic disorder model in the framework of localization theory \cite{Anderson,Lee}. For 2D graphene, a real space order $N$ Kubo method \cite{Roche}
is used to compute the energy-dependent elastic mean free path ($\ell_{e}$),
charge mobilities ($\mu$) and semiclassical conductivities 
($\sigma_{\rm sc}$) in the diffusive regime, before quantum interferences come into play. 
Beyond the diffusive regime,
the energy-dependent localization length ($\xi$) is extracted from the
analysis of the transition from weak to strong localization, following the scaling theory 
phenomenology \cite{Lee}. Quantum transport in GNRs with different 
chiralities (ziz-zag and armchair type) and same disorder 
potential is also investigated within a Landauer approach \cite{Datta}. For GNRs widths in the range $\sim 20-80$ nm 
(within the experimental scope \cite{KimIBM}), it is found that edge effects strongly enhance the impact of disorder, which results in localization
lengths several orders of magnitude smaller than those obtained in
2D graphene for the same disorder strength.

\begin{figure}[h!]
 \centering
 \includegraphics[width=8.5cm]{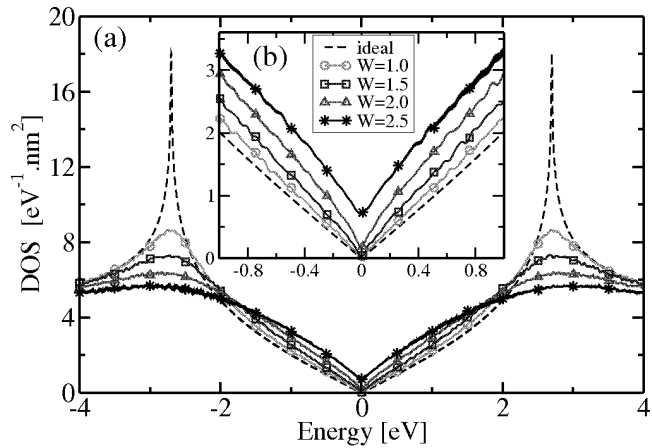}
 \caption{(color online) (a): DoS 
of an ideal (dashed lines) and for 
disordered graphene sheets for several values of 
$W=1, 1.5,2,2.5$ (b): Zoom in the energy area around the CNP.}
 \label{FIG1}
\end{figure}

The low energy electronic properties of 2D graphene are accurately described by the $\pi$-orbital 
tight-binding hamiltonian, which is a first nearest 
neighbor two centers orthogonal $p_{z}$ model, 
with onsite energies $\varepsilon_c=0$ eV for all 
orbitals and hopping term $\gamma_{0}=2.7$ eV. To mimic short range disorder, 
a white noise uncorrelated Anderson type disorder is introduced as a random fluctuation of the onsite 
energies of the hamiltonian ($\varepsilon=\varepsilon_{c}+\delta \varepsilon$). 
The scattering potential can thus be characterized by a single parameter $W$ 
which defines the range of energy variations 
($\delta \varepsilon \in [-W\gamma_{0}/2,+W\gamma_{0}/2]$),
and thus allows to tune the disorder strength. In what follows $W = [0.5,2.5]$ enables the 
exploration of all transport regimes taking place in disordered 2D graphene and GNRs.

In Fig.~\ref{FIG1}, the Density of States (DoS), computed with a Lanczos-type method \cite{Roche}, is reported as a function of disorder strength. The disorder-free DoS (dashed line) 
shows the typical behavior with a linear increase at low energy 
and the presence of two sharp Van Hove singularities at $E=\pm\gamma_{0}$. 
As $W$ is increased, two opposite behaviors are observed. At high energies, 
Van Hove singularities are smoothen whereas close to the charge neutrality point (CNP), 
disorder enhances the DoS in agreement with prior analytical results \cite{AndoJPSJ98} 
(see Fig.~\ref{FIG1}b for a close-up).

To investigate quantum transport in the 2D disordered graphene, 
an efficient real space and order $N$ Kubo method is 
employed \cite{Roche}. In this formalism,
the mean free path $\ell_e(E)$, the semiclassical conductivity
$\sigma_{sc}(E)$ and the charge carrier mobility $\mu(E)$
are deduced
from the energy and time dependence of the diffusion coefficient
$D(E,t)=\langle \Delta R^{2}\rangle(E,t)/t$
(where $\langle \Delta R^{2}\rangle(E,t)$ is the quadratic
spread of random phase wavepackets propagated in the graphene sheet).

\begin{figure}[h!]
 \centering
 \includegraphics[width=8.5cm]{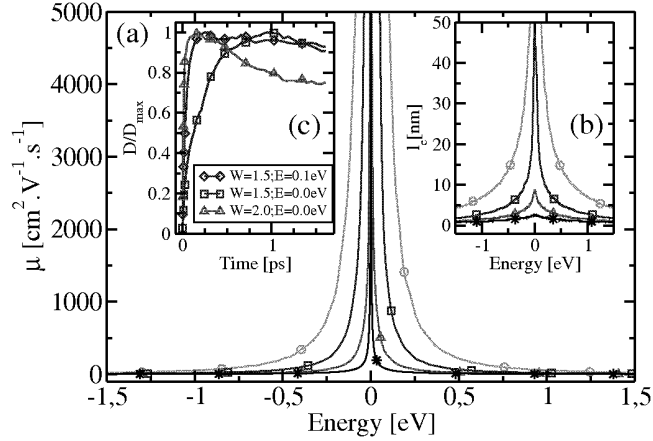}
 \caption{(color online) (a): Energy-dependent 
charge mobility for several values of W.
(b): $\ell_{e}$ for the same disorder strengths.
The legend is the same as in Fig.~\ref{FIG1}.
(c): Diffusion coefficient $D(E,t)$ as a function of 
time for various disorder strengths and Fermi energies.
$D(E,t)$ has been normalized with respect 
to its maximum value $D_{max}(E)$ to allow an easier comparison between the different curves.}
 \label{FIG2}
\end{figure}

In Fig.~\ref{FIG2}c, 
the time dependence of $D(E,t)$ at the CNP and at $E=0.1$ eV 
are reported for two values of $W$ \cite{expl}. 
Different transport regimes follow each other as a function
of the propagation time (or length).
As expected, $D(E,t)$ first scales linearly with $t$ at short times owing to the absence of elastic scattering. 
This linear scaling is followed by a saturation of $D$
at a maximum value $D_{\rm max}(E)$,
that pinpoints the occurrence of a diffusive regime 
for which $D\sim v\ell_{e}$ (with $v$ a group velocity and 
$\ell_{e}$ the elastic mean free path \cite{Roche}). 
As evidenced in Fig.~\ref{FIG2}c, the saturation time decreases 
with increasing disorder strength ($W$) or increasing charge energy (E). At longer times, $D(E,t)$ decreases owing to quantum interferences effects
and localization phenomena \cite{Roche}.
The full energy-dependence of $\ell_{e}$ is given
in Fig.~\ref{FIG2}b for increasing $W$. 

The strong enhancement of $\ell_{e}$ around the CNP
results from the cusp in the DoS, which implies a reduced number of scattering processes. However, $\ell_e(E)$ drops from $180$ nm at $W=1.0$ to $10$ nm 
at $W=2.0$, as a consequence of the increase of the DoS close to CNP. 
In the weak disorder case ($W\in[0.2-0.7]$ not shown here), 
when the DoS at the CNP is almost unchanged with respect to disorder-free graphene case, 
the behavior of $\ell_e(E)$ as a function of $W$ is in good agreement with the Fermi golden rule (FGR)
i.e. $\ell_e(E)\propto 1/W^2$. For higher values of $W$ ($1.0$ to $2.0$) slight deviations to the FGR are expected,
since the weak disorder approximation is not strictly applicable anymore.

Fig.~\ref{FIG2}a shows the corresponding 
charge mobilities deduced from $\mu(E) =\sigma_{\rm sc}(E)/en(E)$, 
where
$\sigma_{\rm sc}=e^{2}\rho(E)v(E)\ell_{e}$ is
the semiclassical conductivity
deduced from the Einstein formula,
$\rho(E)$ is the DoS,
$n(E)$ is the charge density at energy $E$, 
and is $e$ the elementary charge. The energy-dependence of $\mu(E)$ and $\ell_{e}(E)$
are similar, and the sharp increase of $\mu(E)$ in the vicinity of 
CNP is in good qualitative agreement with experimental observations \cite{Graphene2,Graphene3,Graphene4}.

\begin{figure}[h!]
 \centering
 \includegraphics[width=8.5cm]{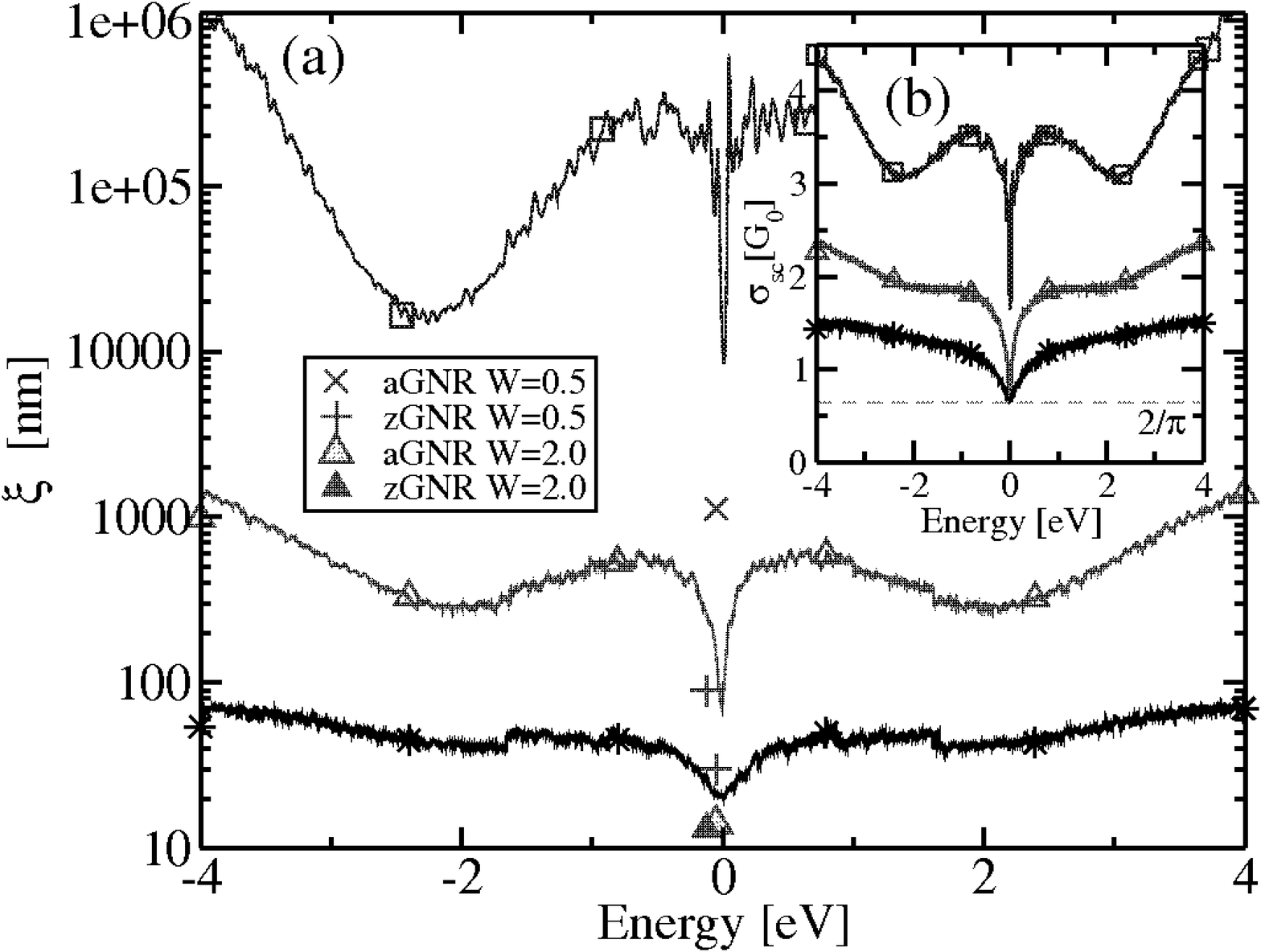}
 \caption{(color online) (a): $\xi(E)$ for three
disorder strengths (same legend as in Fig.~\ref{FIG1}). 
(b): Energy-dependent semiclassical conductivity for the same disorder strengths.}
 \label{FIG3}
\end{figure}

Experiments show that the conductivity (down to a few Kelvin) 
is almost constant close to the CNP, $\sigma(E=0)\sim 3-5 e^{2}/h$, 
and weakly dependent on the value of the charge mobility 
\cite{Graphene1,Graphene2,Graphene3}. On the theoretical side, 
within the SCBA the semiclassical part of the conductivity 
due to short range disorder is found to be 
$\sigma_{\rm sc}=4e^{2}/(h\pi)$ \cite{AndoJPSJ98}. 
By using the Kubo formalism, it was further found that 
$\sigma(E=0)$ strongly depends on the nature of 
the scattering potential (short or long range) \cite{Kubo}. 
Our numerical results are shown in Fig.~\ref{FIG3}b.
$\sigma_{sc}$ clearly remains larger or equal to
$\sigma_{min}=2G_{0}/\pi$ ($G_{0}=2e^{2}/h$), a fact that would be consistent 
with Mott argument \cite{Lee}, although localization 
effects are further observed at all energies (see hereafter). 
Besides, the shape of the energy-dependence of conductivity is 
in perfect agreement with prior analytical results derived for short range disorder within the SCBA \cite{AndoJPSJ98}.

However, the conductivity 
would not be sensitive to localization effects
only in the presence of some decoherence mechanisms 
such as electron-electron scattering of electron-phonon coupling \cite{Lee}.
In contrast, as previously seen in the time-dependence 
of the diffusion coefficient (Fig.~\ref{FIG2}c), 
our zero-temperature calculations evidence the contribution of 
localization effects that develop beyond the diffusive regime. 
The 2D localization length $\xi$ can be evaluated as follows \cite{Lee}:
Whatever the disorder model,
the quantum correction
to the conductivity is expected to scale as
$\Delta\sigma(L) = (G_{0}/\pi)\ln(L/\ell_{e})$,
where $L$ is the length scale associated with the propagation time. The localization length $\xi$ is given by $\Delta\sigma(L=\xi) = \sigma_{sc}$, i.e.
$\xi = \ell_{e} \exp(\pi\sigma_{sc}/G_{0})$. 
Our results are reported in Fig.~\ref{FIG3}a
for several disorder strengths.
The energy-dependence of $\xi$
is mainly dominated by that of $\sigma_{sc}$.
As a result, although $\ell_{e}$ is strongly increasing as
the Fermi level moves towards the CNP (undoped case),
the behavior of $\xi$ shows an opposite trend, with a minimum
value at CNP.

\begin{figure}[h!]
 \centering
 \includegraphics[width=8.5cm]{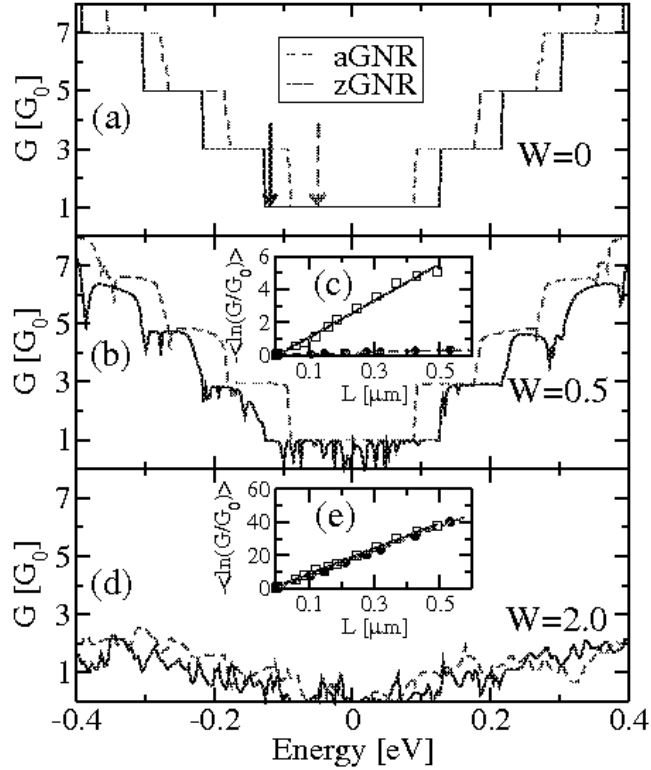}
 \caption{(color online) (a): Conductance for ideal
ziz-zag (solid blue line) and armchair (dashed red line) 
GNRs with 
width of $\sim 20$ nm.
(b): Conductance for a single disorder configuration
of a zig-zag (solid blue line) and an armchair (dashed red line)
GNR with width $\sim 20$ nm and for $W=0.5$.
(c): Configuration averaged (over $\sim 400$ samples) 
normalized conductance as a function of GNR length for 
both zig-zag and armchair GNRs. The solid blue (dashed red) arrow shows the
energy at which the calculations
for the zGNR (aGNR) have been performed.
(d) and (e): Same informations as for (b) and (c) but for a larger disorder strength ($W=2$).}
 \label{FIG4}
\end{figure}

Recently, the possibility to fabricate quasi-1D graphene
nanoribbons has opened new perspectives
for future carbon-based nanoelectronics \cite{Graphene2,KimIBM}. It is thus important
to evaluate the effects of disorder in this situation of lower dimensionality.
Indeed,
the bandstructures of ideal GNRs with
width below $\sim 100$ nm and
well defined edge symmetries (zig-zag or armchair types) are dominated
by confinement effects and Van Hove singularities \cite{Ribbonpure}, similarly to carbon nanotubes \cite{RMP}. 
Zig-zag type GNRs show very peculiar electronic properties 
with wavefunctions sharply localized along the GNRs edges at low energies, 
which will significantly influence their transport properties. 
In contrast, armchair-type GNRs share more similarities with metallic 
nanotubes. Some prior studies have already addressed the impact of homogeneous disorder on
quantum transport in GNRs \cite{White}.

By using a Landauer approach \cite{Datta, Avriller},
$\ell_{e}(E)$ and $\xi(E)$ can be estimated from the scaling analysis
of the average conductance of the GNRs \cite{Avriller}.
In particular, $\xi(E)$ can be accurately extracted from the exponential
decay of the conductance (averaged over $\sim 400$ configurations) versus
length.

In Fig.~\ref{FIG4}, 
the conductance for both types of symmetries are shown in the ideal case and 
for a single disorder configuration
for weak ($W=0.5$) as well as strong disorder 
($W=2$).
The length scaling behaviors of the conductance and the localization lengths
(for both types of GNRs edge symmetries) are respectively
reported in Fig.~\ref{FIG4}c and ~\ref{FIG4}e and in Fig.~\ref{FIG3}.
For disorder as large as $W=2$,
the localization lengths are similar for both types of ribbons,
showing that edge symmetry does not play any role.
In contrast,
$\xi$ is up to two orders of magnitude larger in zig-zag than in armchair GNRs
in the low disorder limit ($W=0.5$), for all values of energy in the plateau around CNP
(some illustrative $\xi$ values at chosen energies are shown in Fig.~\ref{FIG3}).
This can be simply understood as the consequence of the lower dimensionality of
transport in the case of zig-zag symmetry,
driven by more confined wavefunctions \cite{Ribbonpure}.
However,
$\xi$ remains always several orders of magnitude smaller in GNRs than in 2D graphene,
whatever the disorder strength.
Similar results are obtained for GNRs with larger width of
$\simeq 80$ nm (not shown). This points out the predominant role of dimensionality
and edge effects on quantum transport especially in the low disorder limit.

In conclusion, by studying the transport properties in both disordered
2D graphene and GNRs (for short range scattering potential), the impact of 
edge symmetries and transport dimensionality on the transport length scales was outlined. 
Despite the simplicity of the Anderson model,
some of the reported transport features may be generic to other types of disorder such as 
chemical doping, surface functionalization, or topological defects. 
However, the contribution of long range potentials 
(e.g. due to ionized impurities trapped in an oxide) 
deserves further consideration.

BB acknowledges the Institute CARNOT for financial support. We thank the CEA/CCRT supercomputing facilities for providing computational resources.


\begin{thebibliography}{99}
\bibitem{expl}
Periodic boundary conditions are applied.
The convergence is achieved for typical supercell sizes $L_x=320$ nm and $L_y=185$ nm,
which enclose more than 2 millions carbon atoms.
\end{thebibliography}
\end{document}